\documentclass[preprint2]{aastex6} 	

\usepackage{natbib}

\fullcollaborationName{The Friends of AASTeX Collaboration}

\begin{document}

\title{Statistical analysis of acoustic wave parameters near active regions}

\author{M. Cristina Rabello-Soares\altaffilmark{1}}
\affil{Physics Department, Universidade Federal de Minas Gerais, Belo Horizonte, MG 30380, Brazil}

\and 

\author{Richard S. Bogart, Philip H. Scherrer}
\affil{W. W. Hansen Experimental Physics Laboratory, Stanford University, Stanford, CA 94305, USA}

\altaffiltext{1}{cristina@fisica.ufmg.br}

\begin{abstract}

In order to quantify the influence of magnetic fields on acoustic mode
parameters and flows in and around active regions, 
we analyse the differences in the parameters in
magnetically quiet regions 
nearby an active region (which we call `nearby regions'),
compared with those of quiet regions at
the same disc locations for which there are no 
neighboring active regions.
We also compare the mode parameters in active regions 
with those in comparably located quiet regions.
Our analysis is based on ring diagram analysis of all active regions
observed by HMI during almost five years. 
We find that the frequency at which the mode amplitude changes from
attenuation to amplification in the quiet nearby regions is around
4.2 mHz, in contrast to the active regions, for which it is about 5.1 mHz.
This amplitude enhancement (the `acoustic halo effect') 
is as large as that observed in the active regions, and has a very weak dependence on the wave propagation direction.
The mode energy difference in 
nearby regions also changes from a
deficit to an excess at around 4.2 mHz, but averages to zero over all modes. 
The frequency difference in nearby regions
increases with increasing frequency until a point at which the
frequency shifts turn over sharply
as in active regions.
However, this turnover occurs around 4.9 mHz, significantly below the acoustic cutoff frequency.
Inverting the horizontal flow parameters in the direction of the neighboring
active regions, we find flows that are
consistent with a model of the thermal energy flow being blocked directly
below the active region.

\end{abstract}



\section{Introduction} \label{sec:intro}

Solar acoustic waves are generated by turbulent convection 
near the solar surface and propagate in the solar interior. Waves below a certain frequency are trapped in the solar interior and those with a certain frequency and wavelength 
interfere constructively to form standing waves.
Several million normal modes are excited with very small individual amplitudes, and periods around five minutes.
Changes in the wave properties as they propagate through the Sun can reveal
spatial and temporal variations in the interior thermal structure and
dynamics. In particular, the frequencies, amplitudes, and widths of the
modes are modified as they propagate through a sunspot.  
\cite{1987ApJ...319L..27B} were the first to observe reductions in acoustic wave 
power in active regions by up to half of that in magnetically quiet region.
Changes of mode characteristics and flows in active regions have since been
extensively studied using different techniques of local helioseismology,
with the aims of studying the effects of magnetic fields on wave propagation,
generation, and absorption, and also of inferring the subsurface structure
of sunspots
\citep[see, for example,][and references within]{2015ApJ...812...20T, 2013SoPh..287..265B, 2013JPhCS.440a2027B, 2010ARA&A..48..289G, 2008AdSpR..41..846H, 2008JPhCS.118a2084R, 2001ApJ...563..410R}.
However, both the physical mechanisms in play and interpretation of
the measurements are still subject to controversy.

In addition to the power reduction at frequencies below the acoustic cutoff frequency at about 5.5 mHz, there is
a power enhancement of about $40-60$\%
at frequencies between 5.5 and 7.5 mHz 
in the areas surroundings active regions
\citep[][and references therein]{2015LRSP...12....6K}.
This phenomenon, known as the ``acoustic halo'', is observed both in the
photosphere at the edges of active regions \citep{1992ApJ...394L..65B}
and in the chromosphere \citep{1992ApJ...391L.113B}
where it spreads out into the nearby quiet areas.
The increase in size of the acoustic halo with height in the solar atmosphere
enables tomography of the magnetic canopy.

Extraction of energy from acoustic waves incident on an
inclined magnetic field by mode conversion seems to be able to explain
the power reduction 
observed in sunspots \citep[][and references therein]{2015LRSP...12....6K}.
The magnetic fields lower the acoustic cutoff frequency, so that lower
frequency waves can escape into the outer atmosphere.
The mode transformation happens at the critical atmospheric layer
where the magnetic pressure equals the gas pressure.
Mode conversion may also be responsible, at least in part, for the
acoustic halo \citep[][and references therein]{2015LRSP...12....6K, 2015JApA...36...15P, 2016ApJ...817...45R}.
However, one can not exclude other effects
in the complex coupling between oscillations and the propagation medium.

The frequency of global acoustic modes as well as their amplitude and lifetime 
have also been observed to vary with the solar activity cycle
\citep[][and references therein]{2015SSRv..196..137E}.
The frequency variations are correlated with the local surface magnetic
field strength at different latitudes at different phases of the cycle,
following the magnetic butterfly diagram \citep{2008AdSpR..41..846H}.

Unfortunately, the determinations of wave parameters and flows in active regions may be affected by systematic errors due to the influence of the strong magnetic fields on the observational line itself
\citep[e.g.][]{2015SSRv..196..167K}. 
Local helioseismic analysis has always depended upon Doppler-shift
observations of absorption lines that are also sensitive to the
Zeeman effect, in order to obtain simultaneous magnetograms.
As the shape of the lines are affected by strong magnetic fields \citep{2012SoPh..278..217C},
the line shifts obtained in active regions are unreliable
as absolute measurements of the line-of-sight velocity
\citep[e.g.][]{2006ilws.conf...21R}.
To a certain extent this may be true for the helioseismic inferences
derived from them as well, although they depend only on the oscillatory
component, not the absolute value.
Although \cite{2007ApJ...654L.175R}, found that these effects seem to be small,
it is better to use observations from outside areas of strong magnetic field \citep{2013SoPh..282...15C, 2013MNRAS.435.2589C}. Also, because the amplitude of the
oscillations is known to be strongly suppressed for undoubtedly physical reasons in regions of strong magnetic fields, it is not clear when
averaging over finite areas how much of the signal is due to areas
of strong versus weak fields.

In this work, we perform a statistical analysis of the perturbations in the acoustic wave characteristics
in magnetically quiet regions close to active regions compared with those
of quiet regions at the same disc locations but far from any active region.
We have analysed the neighborhoods of each of the active regions observed by 
the Helioseismic and Magnetic Imager (HMI) on board the {\it Solar Dynamics Observatory} \citep{2012SoPh..275..229S}
during a period of almost five years.
We also analyse the perturbations in the parameters obtained at the aforementioned active regions themselves for comparison.
%
In quantifying the effect of the nearby active region on the mode parameters and flows,
we aim to improve our understanding of the 
mechanisms involved in the interaction between acoustic waves and the
strong magnetic fields in sunspots.
We describe the data and our analysis method in section 2, present our results in section 3, and summarize and discuss them in section 4.

\section{Data and Method}

The data used consist of full-disc Dopplergrams obtained at a 45-second cadence by HMI
from 2010 May to 2015 January,
corresponding to Carrington rotations 2097 through 2158.
Small regions of the solar disk with a $5^{\circ}$ diameter were tracked at the Carrington rotation rate for 9.6 hours each
and remapped in solar latitude and longitude using Postel's projection.
Since the horizontal wavelengths of the acoustic modes seen on such
a small scale are much smaller than the solar radius, they can be
treated as plane waves. Hence, the
power spectra calculated by three-dimensional Fourier transforms of the
mapped and tracked data can be fitted to estimate the parameters of
short-wavelength modes. 
This analysis was done by the HMI Ring-Diagram processing pipeline 
for regions with centers separated by $2.5^\circ$ in latitude and
about $2.5^\circ$of arc in longitude, for all regions extending from
disk center to  about $80^\circ$, as described in detail in \cite{2011JPhCS.271a2008B,2011JPhCS.271a2009B}.

The mode parameters were estimated using two different methods for parametric fitting of the ridges in the 3-D power spectrum,
$P(k_x,k_y,\nu)$.
Module `rdfitc' fits a 13-parameter model 
to regions of the power spectrum at a set of selected frequencies within
frequency bands of width $\pm$ 100$\mu$Hz for individual values of the
horizontal wavenumbers $k_x$ and $k_y$
\citep{1999ApJ...525..517B}.
Module `rdfitf' fits only 6 parameters 
to an annular region of the power spectrum at fixed frequencies at selected horizontal wavenumbers $k$ \citep{2000SoPh..192..335H}. 
Both methods use a Lorentzian profile. 
\begin{equation}
P(k_x,k_y,\nu) = \frac{A}{ \Big( \frac{\nu - \nu_d}{\gamma} \Big)^2 + 1}  \,\, S_{\nu} + B(k)
\label{eq_asym}
\end{equation}
where $\nu_d = \nu_0 + k_x U_x + k_y U_y$. 
In the case of rdfitc, there is the addition of an asymmetry term $S$,
defined as in \cite{1998ApJ...505L..51N}, where
$S_{\nu} = S^2 + [1 + S \, (\nu - \nu_d) \,/\, {\gamma}]^2$;
and, for rdfitf, $S=0$, thus $S_{\nu} = 1$.
The amplitude is $A$, $\gamma$ is the half width at half maximum,
$U_x$ and $U_y$ are terms representing the advection due to zonal and meridional motions with respect to the tracking rate, respectively,
and $B(k)$ is the background term.

The six fitted parameters, $p_j$, in rdfitf are: $a_e, \gamma_e, \nu_0, U_x, U_y$, and one value for the background term,
where: 
$\gamma = \gamma_c = \exp(\gamma_e)/2$ and $A=A_c=2\,\exp(a_e - \gamma_e)$.
In this method, the spectrum is filtered, prior to the fitting, to minimize any anisotropy in ring power 
and simultaneously has its resolution decreased to increase processing speed \citep{2000SoPh..192..335H}.
The thirteen fitted parameters ($p_j$, $j=1,13$) in rdfitc are: $A_0, A_1, A_2, A_3, c, p, U_x, U_y, \gamma_c, \gamma_k, S$, and two background terms, where:
$A = \exp[ A_0 + A_1(k-k_0) + A_2 (k_x/k)^2 + A_3 (k_x k_y / k^2) ]$, 
$\gamma = \gamma_c + \gamma_k (k-k_0)$
and the mean frequency $\nu_0 = c k^p$. 
For convenience, we will divide $A$ in three components: 
$A = A_c \,\, A_{\theta} \,\, A_{k}$, 
where $A_{k} = \exp(A_1)$,    
$A_c = \exp(A_0 + \frac{A_2}{2})$
and
$A_{\theta} = \exp[ \slantfrac{A_2}{2} \cos(2\theta) + \frac{A_3}{2} \sin(2\theta) ]$.

For each tracked region of the solar disk, 
the model parameters $p_j$ are obtained for a set of oscillation modes: $p_j(n,l)$.
Each mode is identified by its order $n$, the number of nodes in the radial (or more correctly vertical) direction
and
an effective degree $l$ given by: $l(l+1) = k^2 R^2_{\odot}$,
which 
need not be an integer, since the modes are analysed as plane waves.
Modes are determined with $n$ in the range $0\sbond 3$ for rdfitf and $0\sbond 4$
for rdfitc, 
and with $l$ in the range 400 to 1700 (to 2600 for $f$ modes).

To study how the acoustic waves are modified as they propagate through a sunspot,
we analyze the mode parameters in a magnetically quiet region close to an active region.
The level of magnetic activity in each tracked region is given by the Magnetic Activity Index (MAI) which is also calculated by the HMI processing pipeline.
Briefly, 
using the same mappings and temporal and spatial apodizations as the
tracked Doppler data,
the absolute values of all pixels in corresponding HMI line-of-sight magnetograms with an absolute value greater than 50 G 
($|B_z| > 50$ G) are averaged, with the magnetograms sampled once every
48 minutes (i.e., about 12 magnetograms during the 9.6 hr tracking time)
\citep{2011JPhCS.271a2008B,2011JPhCS.271a2009B}.
We select all quiet patches with MAI $< 5$ G in our data set that have at least one large magnetic active region with MAI $> 100$ G in their vicinity.
These active region patches will be referred to as target tiles ($T$).
As the influence of the active region drops rapidly with distance \citep[for example,][]{2004SoPh..225..213N}, 
we consider only tiles not more than $8^{\circ}$ away from the center of the target tile,
making a total of 36 neighboring tiles
(the nearest non-overlapping tiles and the overlapping ones).
In general, there are several tiles with MAI $> 5$ G in the vicinity of a target tile.
We define the angle between the line from the center of the target tile to the nearby active region and the x-axis (longitude axis),
$\theta_{ar}$,
as the average of the angles of all tiles with MAI $> 5$ G in the vicinity of the target tile 
weighted by their MAI and their distance from the target.
We define also the distance from the target tile to the nearby active region (center to center)
as the average of the distances to all tiles with MAI $> 5$ G in the neighborhood of the target tile
weighted by their MAI. 
For all target tiles analysed here, this distance is on average 6.71 $\pm$ 0.01 degrees.

Observed regions of the solar disc located away from disk center 
have a larger geometric foreshortening and a lower effective spatial resolution, along with different resolution in the zonal and
meridional directions.
In order to avoid the systematic variations that these purely geometric
effects have on the mode parameter determinations,
for each target tile we select a set of 2 to 6 quiet tiles with no nearby active region (i.e., all patches in their vicinity have MAI $<$ 5 G as itself) at the same latitude and Stonyhurst longitude as the target.
We refer to these as the comparison set, $C$.
For a given target patch $i$, the fitted parameters obtained for its comparison set 
are averaged for each mode and subtracted from the fitted parameters obtained for the target patch:
$\Delta p_j(n,l,i) = p_j^T(n,l,i) - \langle p_j^C(n,l,i)\rangle$.
The relative difference of each fitted parameter ($p_j$) for each target tile $i$ is given by:
$\slantfrac{\Delta p_j}{p_j}(n,l,i) = [ p_j^T(n,l,i) - \langle p_j^C(n,l,i)\rangle ] / \langle p_j^C(n,l,i)\rangle$.
There are approximately twice as many targets with $\theta_{ar}$ in the north-south direction ($\pm20^{\circ}$)
as in the east-west direction due to the tracking.
Hence, the parameter differences are averaged over 
ten-degree intervals 
in $\theta_{ar}$ for each mode ($n,l$),
removing outliers greater than 3$\sigma$; these averages,
$\langle \Delta p_j (n,l,\theta_{ar})\rangle$,
are then averaged over all $\theta_{ar}$ for each mode to form the
parameter differences $\langle \Delta p_j (n,l)\rangle$
or
$\langle \slantfrac{\Delta p_j}{p_j}(n,l)\rangle$. These parameter differences will be referred to as the nearby variations (nb).

For control purposes, we arbitrarily choose one of the tiles in the comparison set for each target and 
assume it is the target tile instead. We then repeat the calculation above, 
subtracting the mean of the comparison set (excluding the randomly chosen one) from the fake target,
and averaged over $\theta_{ar}$ intervals (using the value of the real target).
We call this the control set.

We also calculate the difference in the fitted parameters 
obtained for an active tile with MAI between 110 and 120 G and its comparison set, C (as described above).
The range in MAI was (arbitrarily) chosen to be associated with a sunspot.
The parameter differences are then averaged over all tiles, $\langle \Delta p_j(n,l) \rangle$.	
They will be referred to as active region variations (ar) and analyzed in comparison with the nearby variations in the next section.

\section{Results}

\subsection{Amplitude}

\begin{figure*}
\includegraphics[width=1.\textwidth]
{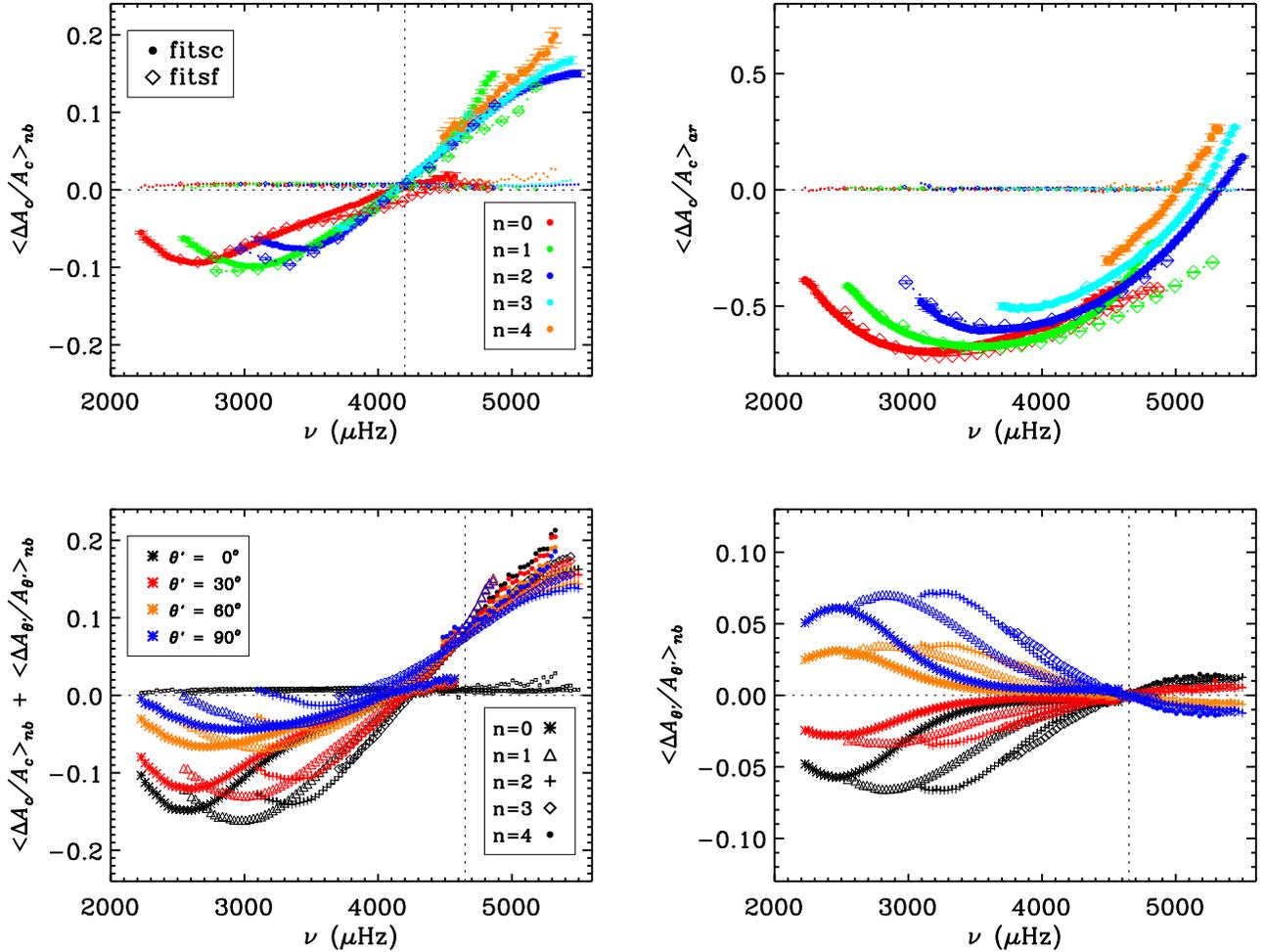}
\caption{\label{amp_near}
Top panels:
The nearby (top left panel) and active region (top right panel) relative variation of the mean power in the ring, $A_c$,
obtained using rdfitc (full circles) and rdfitf (open diamonds and dotted line)
for $f$ (red), $p_1$ (green), $p_2$ (blue), $p_3$ (cyan), and $p_4$ (salmon) modes.
The very small symbols (full circles and diamonds) on both top panels, close to zero, are the results for the control set.
The error bars represent the standard error of the mean which are very small.
Bottom panels:
The nearby relative variation of the maximum power in the ring, $A_c A_{\theta'}$ (bottom left panel) 
and
of the anisotropic component of the amplitude, $A_{\theta'} $(bottom right panel)
for four different directions:
$\theta'=0\degr$ (in the direction of the nearby active region in black),
$\theta'=30\degr$ (red), $\theta'=60\degr$ (salmon), and $\theta'=90\degr$ (perpendicular to it in blue)
obtained using rdfitc 
for $f$ (stars), $p_1$ (triangles), $p_2$ (crosses), $p_3$ (diamonds), and $p_4$ (full circles) modes.
The small black squares close to zero on the bottom left panel are the results for the control set for $\theta'=0$.
The $y$ range is different in the panels.
}
\end{figure*}

Figure \ref{amp_near} (top right panel)
shows the active region relative variation of the mean power in the ring given by $A_c$, 
$\langle \slantfrac{\Delta A_c}{A_c}(n,l)\rangle_{ar}$,
obtained by rdfitc (full circles) and rdfitf (open diamonds and dotted line).
As it is well known, the attenuation is largest around 3 mHz \citep[e.g.][]{2015LRSP...12....6K}.
The measured oscillation power is reduced by $40-70$\% 
for modes with frequency smaller than $\sim$4.8 mHz,  
with some dependence on mode order.
In fact, for frequencies higher than $\sim$3.5 mHz, the power reduction decreases as frequency increases.
It is zero around 5 to 5.3 mHz (depending on mode order) as determined using rdfitc.
At higher frequencies, we start to see an enhancement in the oscillation power due to the acoustic halo effect.
The results of the two fitting methods agree well, except for $f$ and $p_1$ modes with frequency larger than $\sim$4.3 mHz.
As mention in Section 2, rdfitf fits less modes and only until $n=3$.

Figure \ref{amp_near} (top left panel)
shows the nearby relative variation of the mean power in the ring, 
$\langle \slantfrac{\Delta A_c}{A_c}(n,l)\rangle_{nb}$.
It is also mainly a function of frequency.
Mode amplitudes are still attenuated by as much as 10\% around 3 mHz.
It has been observed by several authors that the mode amplitude decreases linearly with the magnetic field strength
\citep[e.g.][]{2008JPhCS.118a2084R,2013SoPh..287..265B},
so the observed maximum nearby attenuation corresponds to active region variation for a region with MAI$\approx$10 G.
The observed maximum nearby attenuation for $f$ and $p_1$ modes happens at lower frequencies than 
the active region variation.
Also, 
the power enhancement happens at a much lower frequency, around 4.2 mHz,
and it is almost as large as in the magnetically active patch (around 20\% at $\sim$5.3 mHz).
The $f$ modes are not amplified for frequencies higher than $\sim$4.2 mHz, 
its variation is close to zero for both fitting methods.
Although, there are some differences between rdfitc (full circles) and rdfitf (open diamonds) results,
there is an overall agreement between them.
The variations given by the control set (small symbols in both upper panels) are much smaller than the target variation.

Simulations by \cite{2015ApJ...801...27R, 2016ApJ...817...45R} support the theory that the acoustic halo enhancement 
is caused by MHD mode conversion through regions of moderate and inclined magnetic fields as proposed by \cite{2009A&A...506L...5K}.
Based on their sunspot model \citep{2015ApJ...801...27R}, 
the distance from the nearby active region is such that 
the quiet regions analysed here have
magnetic field inclinations varying from 65 to 90 degrees (i.e., horizontal),
which would imply a small MAI and
might account for the power enhancement at lower frequencies.

\begin{figure}
\includegraphics[width=0.45\textwidth]
{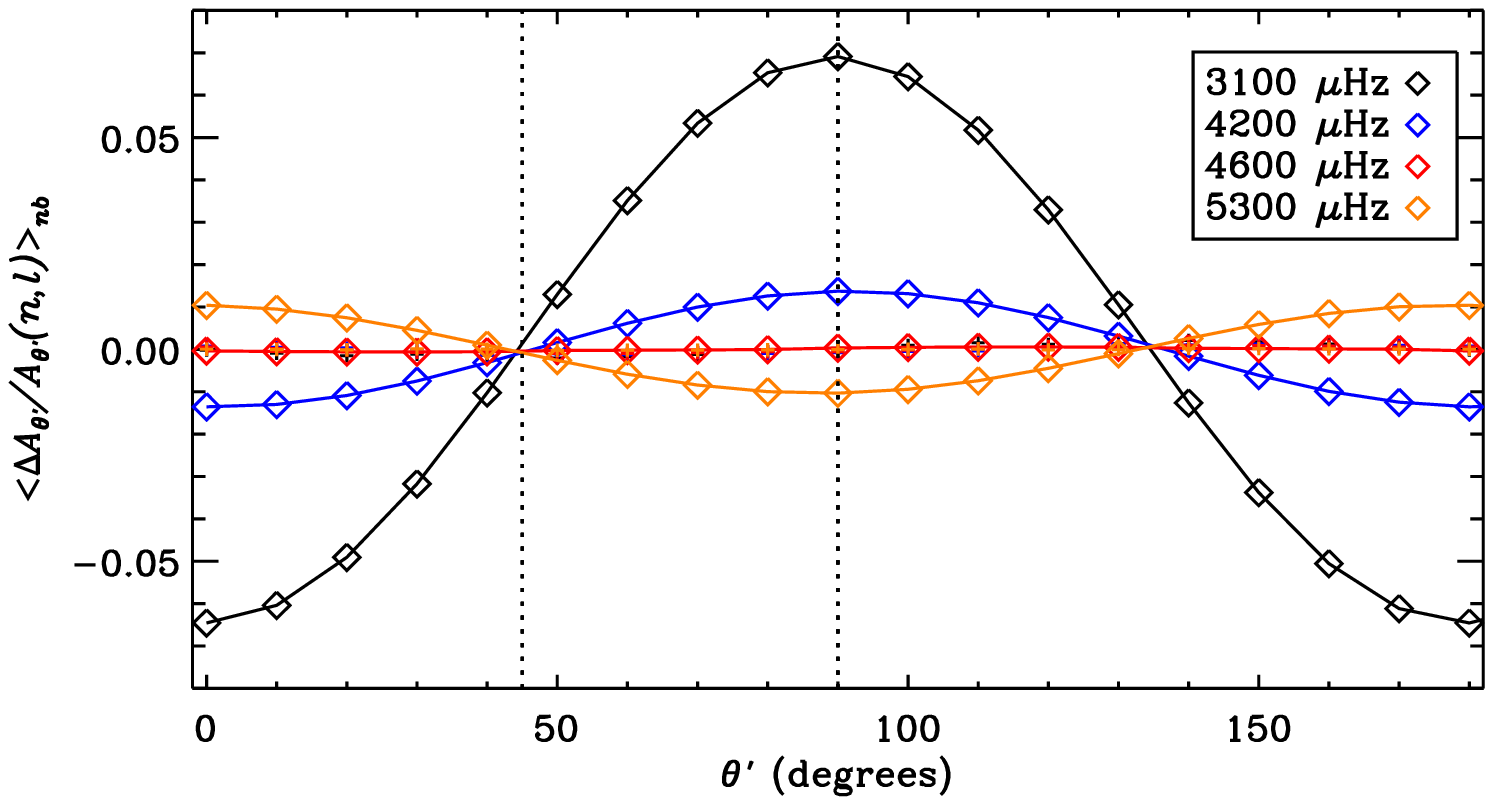}
\caption{\label{atheta_fig2}
Nearby variation of the power anisotropy, 
$\langle \slantfrac{\Delta A_{\theta'}}{A_{\theta'}}(n,l) \rangle_{nb}$ 
for modes with $n=2$ and four different frequencies: 3100 (black), 4200 (blue), 4600 (red), and 5300 $\mu$Hz (orange).
For modes with frequency smaller than 4.6 mHz, it varies as $\cos(2\theta'+\pi)$.
The results for the control set are very close to zero and barely visible (small crosses).
The vertical lines are at 45$\degr$ and 90$\degr$.
}
\end{figure}

The rdfitc method estimates also the anisotropy in the power, $A_{\theta}$ (Section 2).
It is more convenient to express $A_{\theta}$ in relation to 
$\theta_{ar}$ instead of solar latitude and longitude: 
$A_{\theta'} = \exp\{ \slantfrac{A_2}{2} \cos[2(\theta'+\theta_{ar})] + \frac{A_3}{2} \sin[2(\theta'-\theta_{ar})] \}$,
where $\theta' = \theta - \theta_{ar}$.
Figure \ref{atheta_fig2} shows examples of 
$\langle \slantfrac{\Delta A_{\theta'}}{A_{\theta'}}(n,l) \rangle_{nb}$ 
for $p_2$ modes with four different frequencies.
For modes with frequency smaller than 4.6 mHz, 
waves propagating in the direction of the nearby active region 
($\theta'=0^\circ$) 
are attenuated in relation to perpendicular to it ($\theta'=90^\circ$),
as one would expect.
However, 
modes with frequency larger than 4.6 mHz have the opposite behavior
and
modes with frequency close to 4.6 mHz are isotropic.
This is true for all $n$'s as can be seen in 
Figure \ref{amp_near} (bottom right panel) 
which shows $\langle \slantfrac{\Delta A_{\theta'}}{A_{\theta'}}(n,l) \rangle_{nb}$ 
for four propagation directions.
For a given frequency, the anisotropy is larger for modes with a larger mode order $n$ (or a smaller degree $l$).
The anisotropy at higher frequencies is a small effect and it tends to be a
constant value for frequencies higher than 5 mHz where it varies between $\pm$1.2\%.  
Modes with $n=1$ and $\nu \thickapprox 2.8$ mHz or $n=2$ and $\nu \thickapprox 3.2$ mHz 
present the largest anisotropy, varying from $-6.8$\% for $\theta'=0\degr$ to 7\% for $\theta'=90\degr$.

These anisotropic variations are of the same order of magnitude as the attenuation in the constant term (top left panel).
The total amplitude variation is shown in Figure \ref{amp_near} (bottom left panel) by adding the constant term in the amplitude with the azimuthal variation (top left and bottom right panels, respectively).
For $p$ modes with frequency larger than 4.2 mHz, there is an overall increase in mode amplitude due to 
the presence of the nearby active 
and this has very little dependence 
with the propagation direction in relation to the nearby active region. 
Modes with smaller frequencies are attenuated and present a strong anisotropy. 
Those propagating perpendicular to the nearby active 
region have the smaller attenuation, the largest is 4.5\% for $f$ modes with frequency close to 2.9 mHz. 
While modes propagating in the direction of the nearby active can be attenuated by as much as 16\% 
(for $p_1$ modes with $\nu \thickapprox 3$ mHz).

\cite{2011SoPh..268..293C} compared
their numerical simulation of the propagation of  
wave packets through a sunspot model
with observations of active region NOAA 9787 using MDI/SOHO data.
We compared our results in Figure \ref{amp_near} bottom right panel (black points) with
their power spectra determination \citep[Figure~9 in][]{2011SoPh..268..293C} 
for a point on the solar surface 60 Mm behind the sunspot and another at a 45$\degr$ angle and at 85 Mm distance.
The later is used as a quiet-Sun reference, as suggested by them.  
Our results seem to agree.
However, they see an attenuation $\sim$15 times larger than ours (e.g., they observe a $\sim$88\% power reduction for $f$ modes with frequency close to 2.5 mHz, while here it is only 6\%).
At least part of this difference is due to the fact that
our determination is for regions with a 60 Mm diameter which is, on average, 80 Mm away from the sunspot
in comparison to theirs which is a very small region 60 Mm away.

\subsection{Linewidth} 

\begin{figure*}
\includegraphics[width=1.\textwidth]
{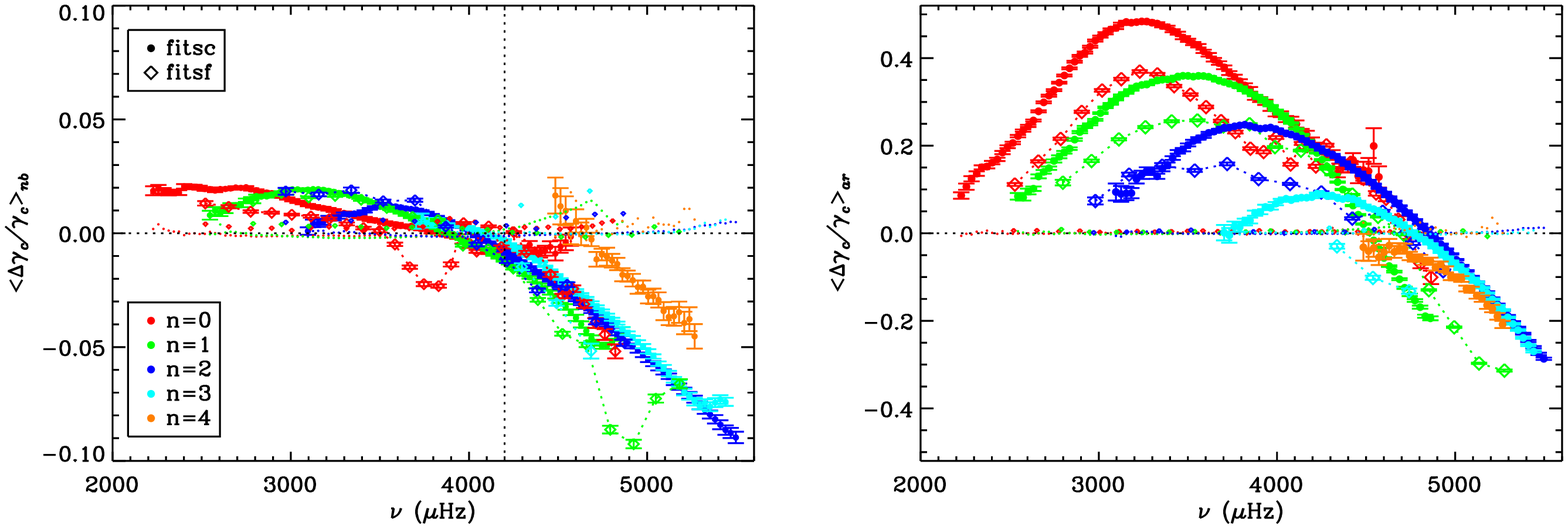}
\caption{\label{wdt_near}
The nearby (left panel) and active region (right panel) relative variation of the half width at half maximum of the peak, $\gamma_c$,
obtained using rdfitc (full circles) and rdfitf (open diamonds and dotted line)
for $f$ (red), $p_1$ (green), $p_2$ (blue), $p_3$ (cyan), and $p_4$ (salmon) modes.
The small symbols (full circles and diamonds) on both top panels, close to zero, are the results for the control set.
The error bars indicate the standard error of the mean.
}
\end{figure*}

The linewidth of a peak in the power spectrum 
is inversely proportional to the lifetime of the mode and contains information of the mode damping.
Figure \ref{wdt_near} (right panel) shows the active region variation of half width at half maximum of power in ring,
$\langle \slantfrac{\Delta \gamma_c}{\gamma_c} (n,l)\rangle_{ar}$.
Although they give similar results, rdfitf variation is
systematically lower (by 10\% on average) than rdfitc.
The active region linewidth variation increases as frequency decreases 
below $\sim$4.8 mHz
until a certain frequency which varies with the mode order.
These agrees with previous results \citep[e.g.,][]{2001ApJ...563..410R,2008JPhCS.118a2084R}
and also with solar cycle variations \citep[e.g.,][and references within]{1990LNP...367..189P,2000ApJ...531.1094K}.

Figure \ref{wdt_near} (left panel) shows the linewidth nearby variation,
$\langle \slantfrac{\Delta \gamma_c}{\gamma_c} (n,l)\rangle_{nb}$.
The variation is also positive (but only by 2\% or less) for frequencies smaller than $\sim$4.2 mHz.
%
%
On the other hand, the linewidth nearby variation for modes with frequency larger than 4.2 mHz decreases by as much as 10\%. 
This decrease is mainly a function of frequency, except for $p_4$ modes that have a similar dependence with frequency (i.e. a similar slope) as the other modes but are displaced towards less negative values.
The results using rdfitc (full circles) and rdfitf (open diamonds) agree in general, but there are some differences. 
The results for rdfitf have a peculiar behavior 
with valleys for $f$ and $p_1$ modes with degree around 1500
and frequencies around 3.8 mHz and 4.9 mHz respectively.
This seems not to be an artifact of the fitting method, since
it is not present in the variations given by the control set (small symbols) which are very small;
and we do not have an explanation at present.

%
%
Although the nearby variation for $f$ modes with frequency larger than 4.2 mHz is similar to the $p$ modes
using rdfitf,
it is close to zero for rdfitc.

\begin{figure*}
\includegraphics[width=1.\textwidth]
{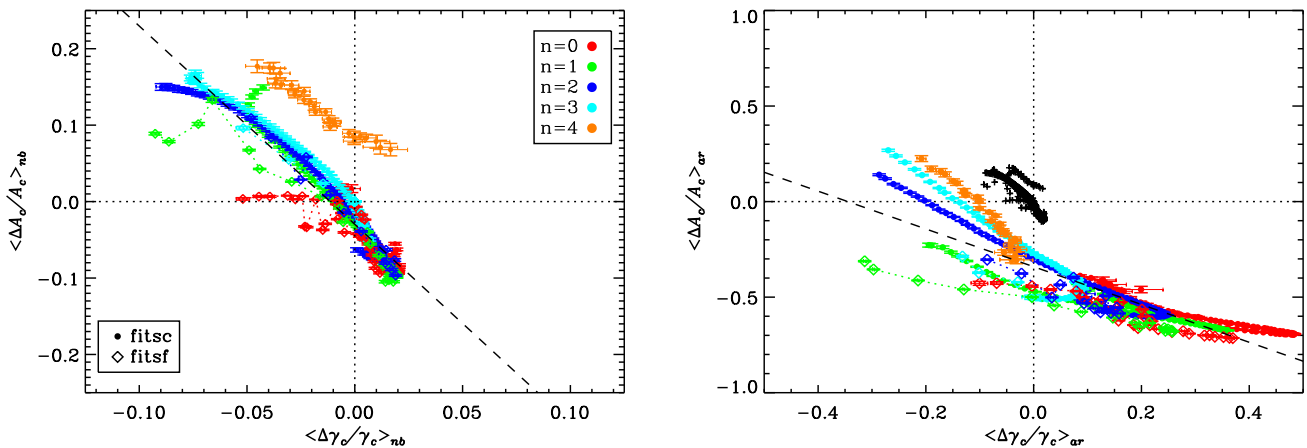}
\caption{\label{wdt_amp}
Relative nearby (left) and active region (right) amplitude variation as a function of the 
width variation 
obtained using rdfitc (full circles) and rdfitf (open diamonds and dotted line)
for $f$ (red), $p_1$ (green), $p_2$ (blue), $p_3$ (cyan), $p_4$ (salmon) modes.
For comparison purpose, 
the data in the left panel are plotted on the right panel as black crosses.
The dashed line in both panels is the linear fit to the data.
The vertical and horizontal error bars show the standard error of the mean.
}
\end{figure*}

Figure \ref{wdt_amp} (right panel) shows the active region amplitude variation versus the width variation.
Fitting a straight line to the data, it will have a slope of $-1$ and an y intercept equal to $-0.34$ (dashed line);
implying that the mode kinetic energy is lower by roughly the same amount for all modes in an active region.
While the nearby variation (left panel) has a slope of $-2.6$ and an y intercept equal to $-0.03$ (dashed line).
The points in the left panel are also plotted in the right panel (black crosses) for comparison.
As in the linewidth nearby variation plot, $p_4$ modes show a distinctive variation for other $p$ modes.
The peculiar behavior of the variation for $f$ and $p_1$ modes using rdfitf is seen here.

\subsection{Energy} 

\begin{figure*}
\includegraphics[width=1.\textwidth]
{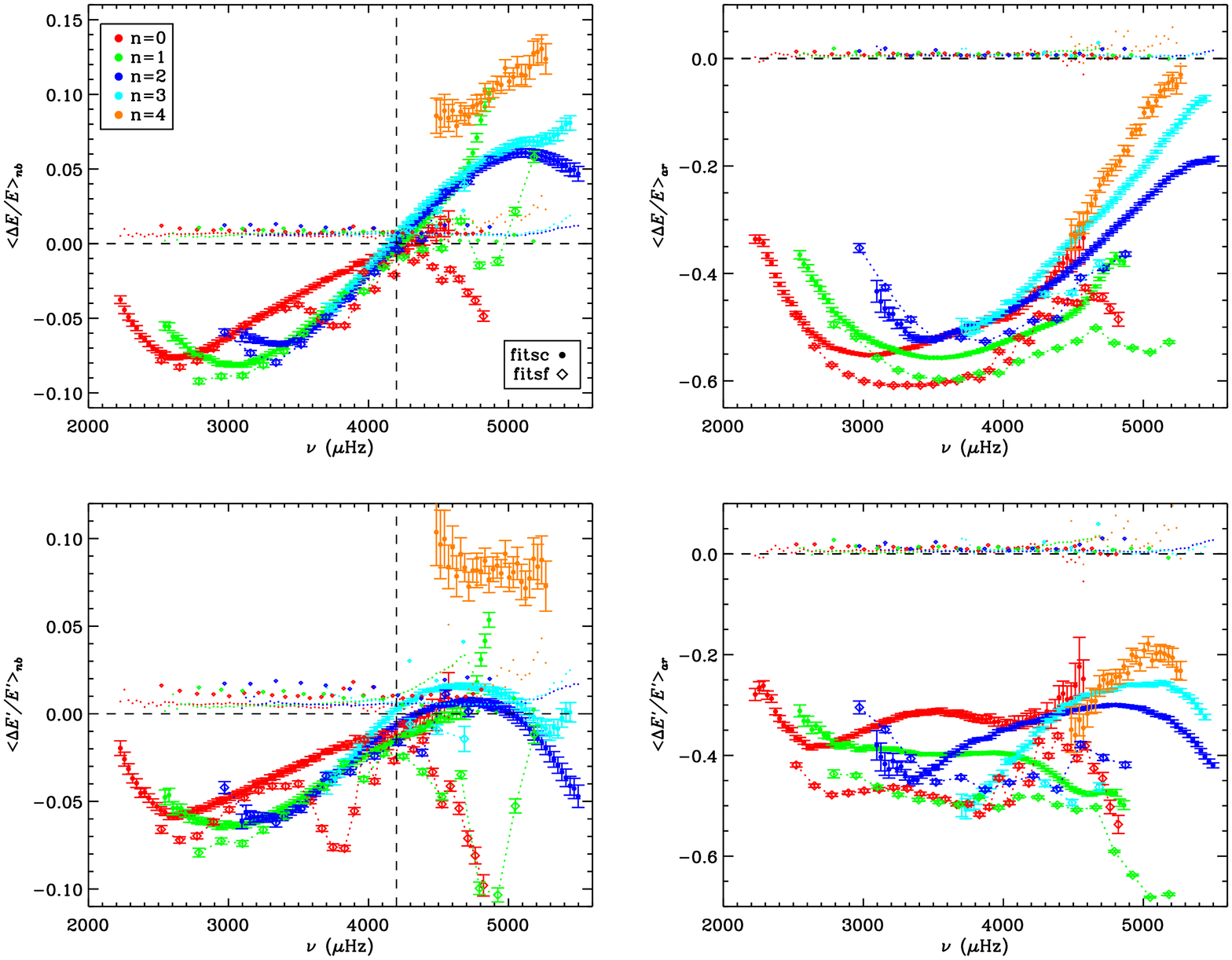}
\caption{\label{ener_near}
Top panels:
The nearby (top left panel) and active region (top right panel) relative variation of the mode energy
($E \propto A_c \, \gamma_c$).
Bottom panels:
The nearby (bottom left panel) and active region (bottom right panel) relative variation of the mode energy rate
($\dot{E} \propto A_c \, \gamma_c^2$).
All panels show the variations 
obtained using rdfitc (full circles) and rdfitf (open diamonds and dotted line)
for $f$ (red), $p_1$ (green), $p_2$ (blue), $p_3$ (cyan), and $p_4$ (salmon) modes.
The small symbols (full circles and diamonds) on all panels, close to zero, are the results for the control set.
The error bars correspond to the standard error of the mean.
}
\end{figure*}

The surface velocity power of the mode
is given by the area under the peak in the power spectrum
which can be approximated by the product of peak height and width.
It is proportional to the mode mean square surface (radial) velocity averaged over time and over the solar surface.
Thus it is also proportional to the mode energy (given by the mode kinetic energy).
Figure \ref{ener_near} (top right panel) shows the mode energy active region variation,
$\langle \slantfrac{\Delta E}{E} (n,l)\rangle_{ar}$
where $E \propto A_c \, \gamma_c$.
There is a decrease in the mode energy in the active region
for all modes analyzed here, even close to the cutoff frequency.
Apparently, the increase in the mean power (given by $A_c$) at frequencies higher than 5 mHz obtained by rdfitc
(Figure \ref{amp_near} top right panel) 
is compensated by the decrease in the linewidth (Figure \ref{wdt_near} right panel).
The decrease in mode energy 
is larger at frequencies smaller than $\sim$4.2 mHz, where it is $-50$\% on average, similar to
the variation observed by others \citep[e.g.][and references within]{2001ApJ...563..410R}.

The nearby energy variation (top left panel) still show a decrease in mode excitation at frequencies smaller than 4.2 mHz 
by as much as ~8\%. 
The relative energy decrease is very similar to the amplitude (Figure \ref{amp_near} top left panel), 
since the linewidth increases very little (Figure \ref{wdt_near} left panel).
At higher frequencies, there is a clear increase in the mode energy.
The nearby energy variation is mainly a function of frequency
varying from large negative values around 2.4 mHz towards zero at around 4.2 mHz
and becoming increasingly large until 5 mHz.
There is, however, 
for $f$ and $p_1$ modes with frequency larger than 4.2 mHz,
a large disagreement between the two fitting methods. 
This can also been seen in the active region variation.
As in the linewidth variation, $p_4$ modes have a distinct behavior in the nearby variation. 
Although the variation observed for $p_4$ modes using the control set is larger than for other modes and
increases with frequency (small symbols in the same panel), it accounts only for half of the increase observed:
$\Delta \langle \Delta E/E \rangle_{nb}$ = 0.02 against 0.04.
The linewidth variation observed for $p_4$ modes using the control set is much smaller,
making it difficult to attribute $p_4$ mode distinctive behavior to bad fitting.
The control set variation for other modes is very small but it is systematically displaced by ~0.07
for nearby and active region energy variation.
A very small systematic increase in the amplitude variation is also present in Figure \ref{amp_near}.

Assuming, as in \cite{1988ApJ...334..510L},
that each mode behaves like a damped oscillator, 
the temporal rate at which the energy is pumped in or out the mode,
$\dot{E}$,
is proportional to $A_c \, \gamma_c^2$.
The temporal rate at which energy is removed from the mode at the active region (Figure \ref{ener_near} bottom right panel) 
does not vary strongly with frequency as observed for the other parameters.
It is $-(40 \pm 10)\%$ on average for the modes analyzed here.
\citep{2000ApJ...543..472K} observed a decrease in the energy rate with the increase of solar activity
with no frequency dependence.  
While, at the nearby region, 
energy is been removed from modes with frequency smaller than 4.2 mHz
(Figure \ref{ener_near} bottom left panel);
it is maximum around 3 mHz. 
It is close to zero for modes with larger frequency,
except for $p_4$ modes which seem to have a supply of energy 
(which would contribute to the halo effect). Again the variation observed in the control set is not enough to
explain their behavior.
The results obtained by rdfitf for 
$f$ and $p_1$ modes show a more chaotic behavior, with strong decreases in the energy rate (around 3.8 mHz and 4.8 mHz).
The variation using the control set is systematically displaced ~1\% and is larger at higher frequencies.

\subsection{Asymmetry}

\begin{figure*}
\plottwo{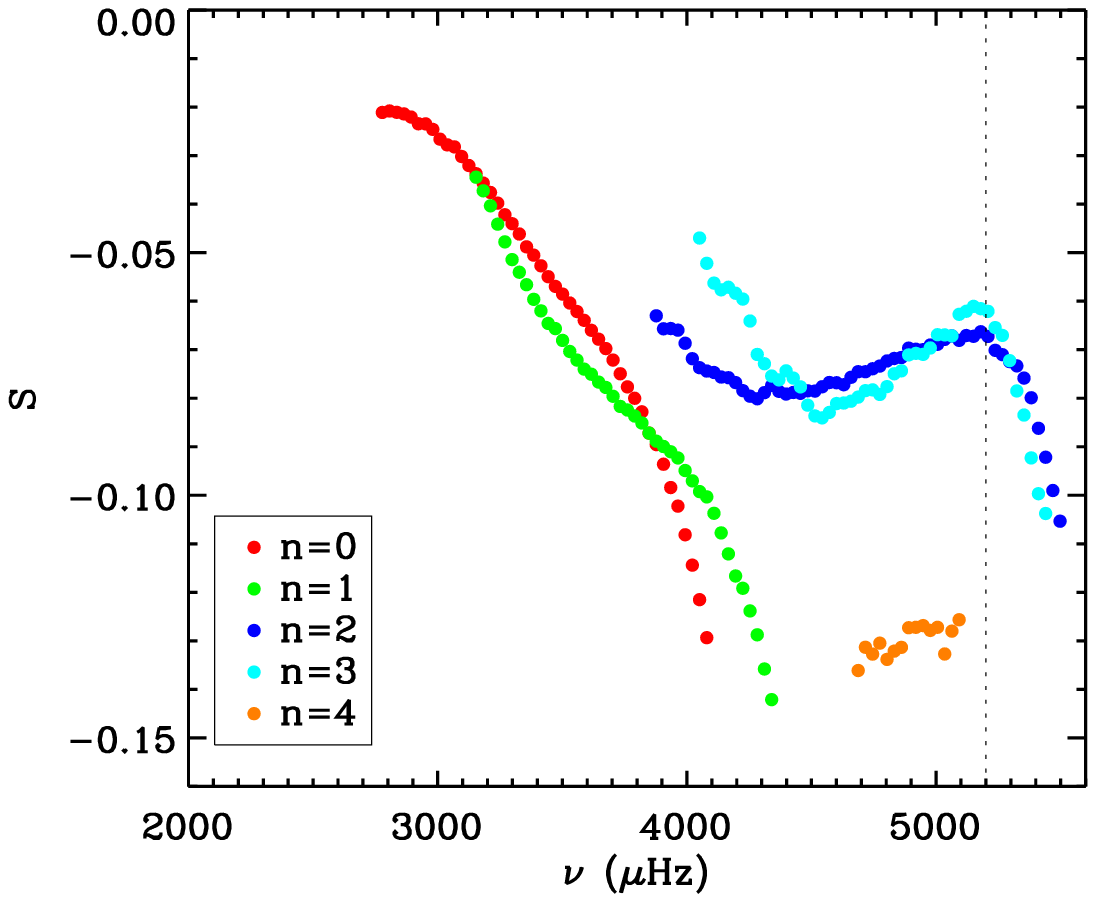}{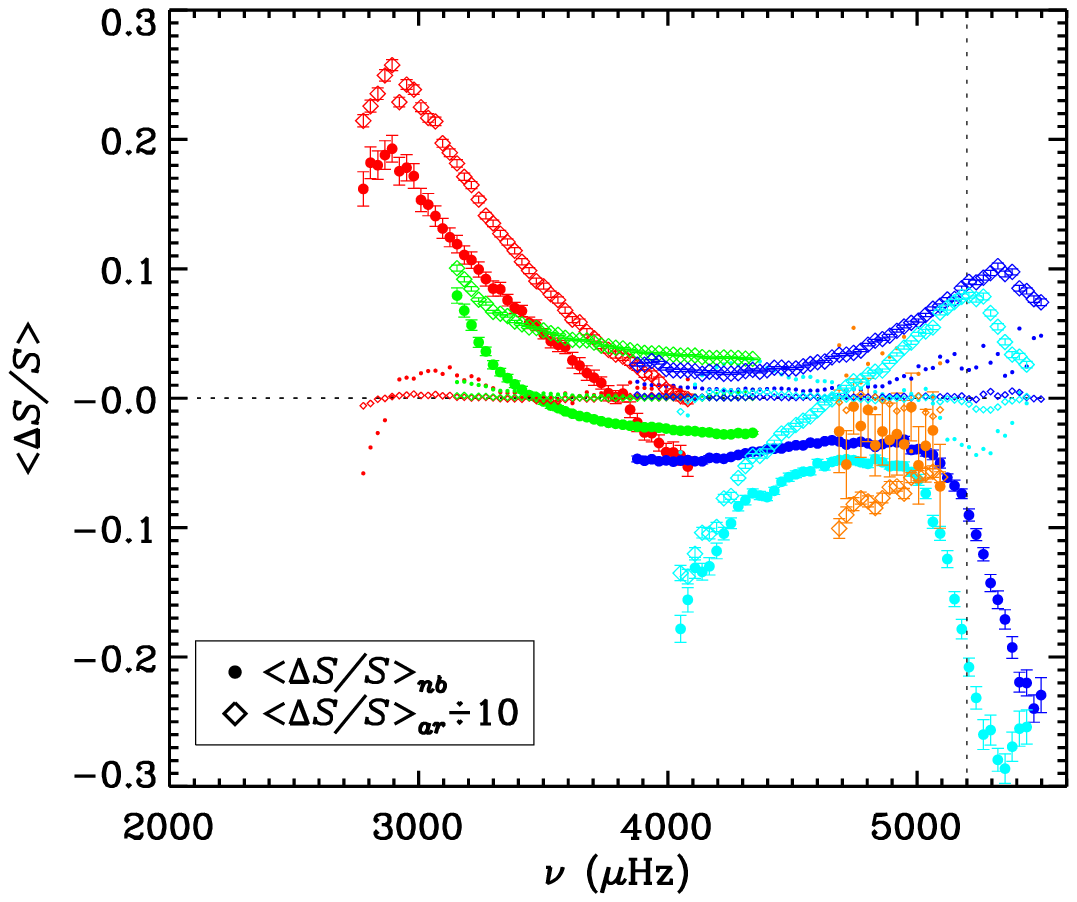}
\caption{\label{asym_near}
Left panel: Observed peak asymmetry, $S$, averaged over all quiet tiles for each mode.
Right panel:
The nearby (full circles) and active region (open diamonds) relative variation of the peak asymmetry obtained using rdfitc.
The small symbols (full circles and diamonds), close to zero, are the results for the control set for the nearby and active region variations.
Their values at low frequency are not small in comparison with the target variation and were excluded from the analysis.
The active region variation was divided by 10 for visualization purposes.
The error bars indicate the standard error of the mean.
Both panels show the results for $f$ (red), $p_1$ (green), $p_2$ (blue), $p_3$ (cyan), and $p_4$ (salmon) modes.
}
\end{figure*}

The asymmetry in the power spectra profile depends on the depth and properties of the acoustic sources responsible for exciting the acoustic modes and the part of the solar background noise that is correlated with the acoustic modes \citep[e.g.][and references within]{1993ApJ...410..829D, 1999ApJ...519..396K, 2006MNRAS.371.1731T}.
The larger the absolute value of the asymmetry parameter $S$, the more asymmetric is the peak. 
Previous studies found that 
the asymmetry is negative (i.e., 
the power in the low-frequency half of the peak is higher than that in the high-frequency part),
as in Figure \ref{asym_near} (left panel).

Figure \ref{asym_near} (right panel) shows the active region and nearby variations of the line asymmetry,
$\langle \slantfrac{\Delta S}{S} (n,l)\rangle_{ar}$ 
(open diamonds and full circles respectively) obtained using rdfitc.
The active region variation is divided, arbitrarily by 10 to make it comparable with the nearby variation at lower frequencies. 
The asymmetry variation at low frequency obtained for the control set (small symbols) is large
and it was excluded from the analysis.
The asymmetry seems to increase with the magnetic activity index \citep[e.g.][]{2001ApJ...563..410R, 2008JPhCS.118a2084R, 2011JPhCS.271a2030S}, 
which suggests that the modes are excited closer to the surface in active regions than in a quiet region
\citep{1999ApJ...519..396K}. 
The asymmetry active region variation is positive for all modes 
except for $p_3$ modes with frequency smaller than 4.6 mHz and all $p_4$ modes observed here
(Fig. 6 right panel).
This also agrees with results for variation with solar cycle, where
\cite{2007ApJ...654.1135J} reported 
a 15\% increase in the asymmetry negative values 
from minimum to maximum solar activity
for low-degree modes ($l < 3$).
The active region variation has a maximum for $f$ modes around 3 mHz 
where it is 2.5 times larger than at a quiet region
and a local maximum around 5.2 mHz for $p_2$ and $p_3$ modes which the asymmetry is 100\% larger.
The nearby variation also has a maximum for $f$ modes around 3 mHz where it is 20\% larger due to the presence of
a nearby active region.
For the $p$ modes, the nearby variation is quite different from the active region one.
Around 5.2 mHz, where there is a local maximum for the active region variation,
there is a minimum (or a maximum decrease in the asymmetry) at the nearby region, by almost 3\%.

\subsection{Frequency}

\begin{figure*}
\includegraphics[width=1.\textwidth]
{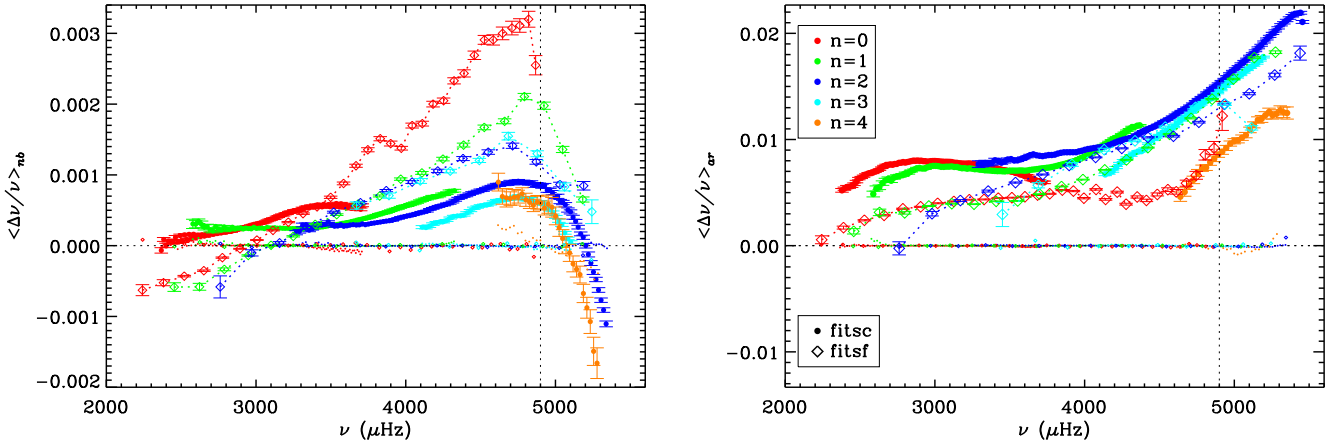}
\caption{\label{freq_near}
The nearby (left panel) and active region (right panel) relative variation of the mode frequency
obtained using rdfitc (full circles) and rdfitf (open diamonds and dotted line)
for $f$ (red), $p_1$ (green), $p_2$ (blue), $p_3$ (cyan), and $p_4$ (salmon) modes.
The small symbols (full circles and diamonds) on both panels, close to zero, are the results for the control set.
The error bars show the standard error of the mean.
}
\end{figure*}

In the presence of strong magnetic fields, it is well observed an increase of the mode frequency which is larger at larger frequencies until near the cutoff frequency where it sharply decreases.
Figure \ref{freq_near} (right panel) shows the active region variation of mode's central frequency given by 
$\nu_0$ in equation (1).
This can be compared with results using $15^{\circ}$ patches by \cite[e.g.][and others]{2001ApJ...563..410R, 2004ApJ...610.1157B, 2008JPhCS.118a2084R}.
The two fitting method show a similar increase with frequency, but the variation is systematically lower for rdfitf.
The nearby frequency variation also increases with frequency by as much as 0.1\% (Figure \ref{freq_near} left panel). 
Although, this is an order of magnitude smaller than the active region variation,
it is significative in comparison with 
the very small control set variation, which is given by the small symbols.
The two fitting methods give different results for the nearby variation.
However, they both show an increase with frequency until around 4.9 mHz,
notably below the acoustic cutoff frequency,
where the variation sharply decreases, becoming negative in the case of rdfitc.

In fact, 
here we do not observe high enough frequency to see the sharp decrease in the active region frequency variation
close to the cutoff frequency where it becomes negative,
which was first observed by \cite{1994SoPh..150..389R} and more recently by \cite{2002ESASP.508...37R, 2008AdSpR..41..846H, 2011JPhCS.271a2029R} among others.
This decrease in the mode frequency at active regions
has been observed also as an increase in the degree (or wavenumber) at a given frequency \citep{2011SoPh..268..349S}.
\cite{1996ApJ...456..399J} using a simple model of a chromospheric atmosphere with a horizontal magnetic field on top
of a simple polytrope
observed this change in the frequency shift sign
and that the frequency where it happens depends on the amount of increase in
magnetic field and temperature of the chrosmosphere.
The fact that 
the sharp decrease observed at the nearby frequency variation happens at a lower frequency than at the active region could be due to a stronger horizontal magnetic field and/or temperature at the chromospere near an active region.

As in the case of active region variation, the nearby frequency shift is mainly a function of frequency
following roughly the inverse of 
inertia of the mode (before the sharp decrease).
This indicates that the main mechanism responsible for the shifts is located very close to the solar surface,
less than 0.5 Mm below the photosphere \citep{2015SSRv..196..137E}.
However, after removing this 'surface term' given by
the frequency variation multiplied by the scaled mode inertia $Q(n,l)$ 
(i.e., normalized by the inertia of a radial mode of the same frequency),
it has been estimated variations in the sound speed below an active varying between 3 and 10 Mm deep 
\citep{2008SoPh..251..439B, 2013SoPh..287..265B},
which coincide in depth and magnitude with variation between solar minimum and maximum \citep{2012ApJ...745..184R}.
\cite{2009SoPh..257...37L} showed that the above sound-speed determination have also a magnetic component
since the solar models used in its determination did not include magnetic fields.
The interpretation of these frequency variations has proven to be quite complicate
and different local helioseismology methods give different results \citep{2009SSRv..144..249G}.

The scaled nearby frequency variation, on the other hand, 
deviate very little from a smooth function of frequency
indicating that the observed frequency variation is due to near-surface effects.
The linear regression slope of the so called `surface term', 
$\langle \slantfrac{\Delta \nu}{\nu}(n,l)\rangle_{nb} * Q(n,l)$,
is 0.43 $\pm$ 0.03 mHz$^{-1}$ for rdfitc and three times larger for rdfitf 
(1.4 $\pm$ 0.1 mHz$^{-1}$), 
in the frequency interval from 2.4 to 4.9 mHz.
Note that the nearby frequency variation obtained by rdfitf is negative for frequencies smaller than 3 mHz
(Figure \ref{freq_near} left panel). 
The corresponding standard deviation to the straight line 
is 0.1 and 0.2 for rdfitc and rdfitf respectively.
%
On the other hand,
for the active region variation, 
the standard deviation from a linear regression in frequency is 
much larger, it 
is 4 and 2 for rdfitc and rdfitf respectively.

\subsection{Flow}

\begin{figure*}
\includegraphics[width=1.\textwidth]
{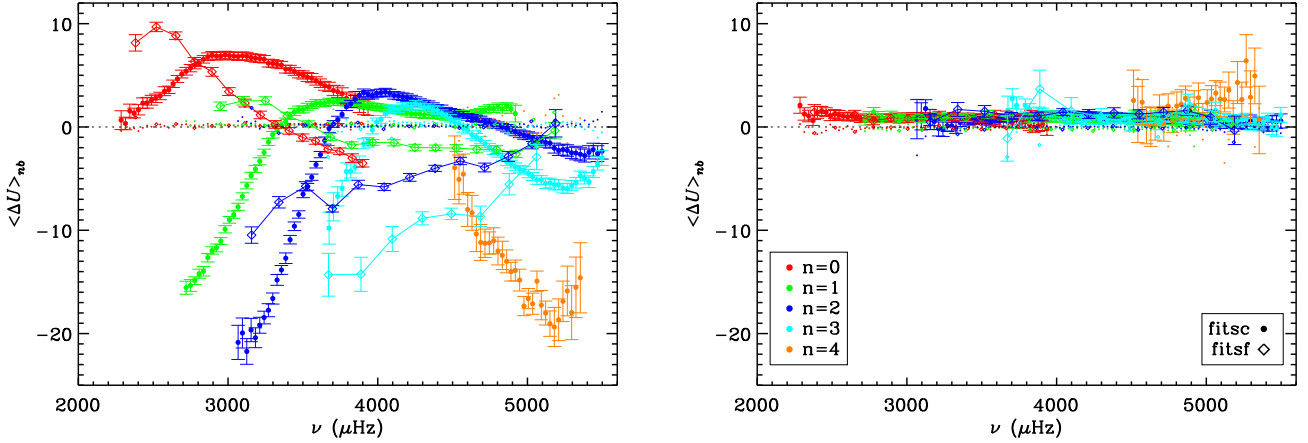}
\caption{\label{fitted_flow}
Nearby flow variation in the direction of the nearby active region (left)
and perpendicular to it (right)
obtained using rdfitc (circles) and rdfitf (diamonds)
for $f$ (red), $p_1$ (green), $p_2$ (blue), $p_3$ (cyan), $p_4$ (salmon) modes.
The small symbols are the variations given by the control set.
The error bars represent the standard error of the mean.
}
\end{figure*}

We will change the reference system for 
the flow from solar longitude-latitude 
given by zonal and meridional flows, $U_x(n,l)$ and $U_y(n,l)$ (Equation 1), 
to parallel-perpendicular to the nearby active region,
by rotating the coordinate system for each target tile by its $\theta_{ar}$ (defined in Section 2).
The flow differences, $\langle \Delta U_{\shortparallel \,,\, \perp} (n,l,\theta_{ar})\rangle_{nb}$,
are averaged over all $\theta_{ar}$, 
giving
$\langle \Delta U_{\shortparallel} (n,l)\rangle_{nb}$
and $\langle \Delta U_{\perp} (n,l)\rangle_{nb}$ 
which are plotted in 
Figure \ref{fitted_flow}.
There is no significant variation in $\langle \Delta U_{\perp} (n,l)\rangle_{nb}$ (right panel).
On the other hand, the variation in $U_{\shortparallel}$ (left panel) 
is much larger than the control set (small symbols).
The variation in the direction of the nearby active region 
obtained by the two fitting methods are different.

\begin{figure}
\includegraphics[width=0.45\textwidth]
{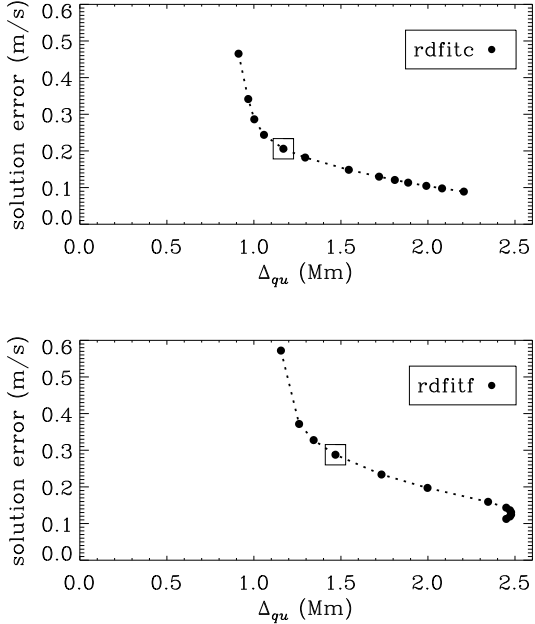}
\caption{\label{flow_trade}
Trade-off diagram at a depth of 3.5 Mm for the flow variations obtained by rdfitc (top) and rdfitf (bottom).
There are 13 parameters from $\mu$ = 0.00001 to 0.01150.
The chosen parameter is shown by a square.
}
\end{figure}

$\langle \Delta U_{\shortparallel} (n,l)\rangle_{nb}$ are inverted 
to infer the variation with depth using the optimally localized average 
\citep[OLA;][]{1968GeoJ...16..169B} technique
as described in \cite{1999ApJ...512..458B}.
This inversion procedure was implemented in module rdvinv of the HMI Ring-Diagram processing 
pipeline \citep{2011JPhCS.271a2008B,2011JPhCS.271a2009B}, which was used in this work.
For the purpose of inversion, the
fitted values of $\langle \Delta U_{\shortparallel} (n,l)\rangle_{nb}$ are interpolated to the nearest
integral value of $l$. 
There is a trade-off parameter in the inversion procedure, $\mu$, determining the balance between
a well localized solution
and a small solution error.
Figure \ref{flow_trade} shows a trade-off diagram at a depth of 3.5 Mm
for the inversion results using the flow variations given by rdfitc (left panel) and rdfitf (right panel)
where
the measurement of how well the solution is localized is given by 
$\Delta_{qu}$ which is the difference between the first and third quartile points of the averaging kernel.
The averaging kernel is defined as $\sum_{n,l} c_{n,l} \, K_{n,l}(r)$
where $K_{n,l}$ are the mode kernels and $c_{n,l}$ are the inversion coefficients.
The best trade-off parameter for the data set obtained by rdfitc and by rdfitf are
$\mu = 0.0005$ and $\mu = 0.0002$ respectively (squares in Figure \ref{flow_trade}).

\begin{figure*}
\includegraphics[width=1.\textwidth]
{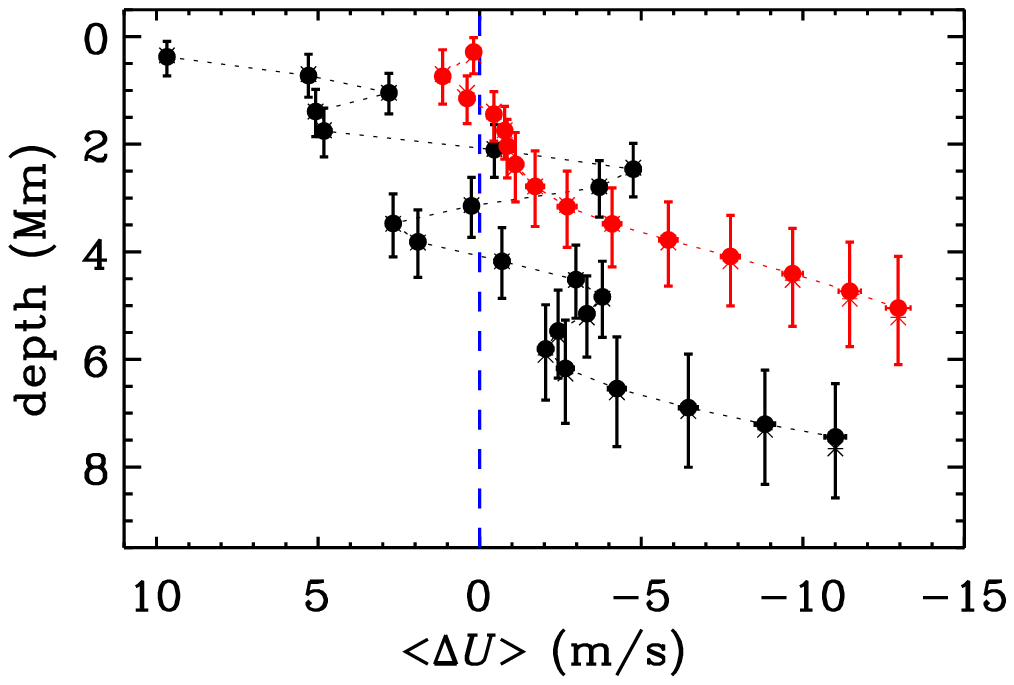}
\caption{\label{flow_sol}
Solution for the nearby flow variation 
in the direction of the nearby active region
obtained for rdfitc (black) and rdfitf (red),
where negative flows indicate flows moving away from the nearby active region (outflows) and
and positive flows moving towards the nearby active region (inflows).
The depth is given by the second quartile point of the averaging kernel (full circles),
while the vertical bars extend between the first and third quartile ponts.
The tentative target radii are shown with a star.
The small horizontal error bars are the one-sigma error in the solution.
}
\end{figure*}

Figure \ref{flow_sol} shows the solution for the parallel flow variation	
obtained for rdfitc (black) and rdfitf (red) data sets.
The inferred flows 
are negative if moving away from the nearby active region (outflows) and
positive if moving towards the nearby active region (inflows). 
The estimated flows are an average over
a quiet region of 60 Mm in diameter with an active region 80 Mm away on average from its center.
Since we are observing a small patch in the solar surface and, hence, fitting only modes with high degree that are trapped 
close to the solar surface, inversions are obtained only until a few Mm.
The uncertainty in 
$\langle \Delta U_{\shortparallel} (n,l)\rangle_{nb}$ 
is given by the error average of a large sample and is very small (Figure \ref{fitted_flow})
and, consequently, the inversion solution errors are also very small.

There is a clear disagreement between the solutions for rdfitc and rdfitf.
rdfitc method fits a much larger number of modes than rdfitf (324 and 64 respectively) in the same power spectra range.
The fewer modes obtained by rdfitf 
are not enough to obtain solutions as deep and as well localized as rdfitc.
On the other hand,
the solution for rdfitc presents an oscillatory behavior 
which is likely due to 
correlation between the errors in the solution at different depths.
The errors in the solution at positions close to each other are generally correlated because they have been derived from the
same mode set and this can introduce features in the solution as described by 
\cite{1996MNRAS.281.1385H}.

Despite the disagreement between the fitting methods, they both observe
a flow towards the active region at depths of, approximately, 0.5 to 1.5 Mm. 
Using rdfitc, this flow is around 5m~s$^{-1}$, while for rdfitf, it is only 0.5 m~s$^{-1}$.
Both methods observed a region of zero horizontal flows around 2 Mm followed by a region of outward flows below.
At a depth of 5 Mm, rdfitc estimates a flow of 3 m~s$^{-1}$ and rdfitf, 10 m~s$^{-1}$.
According to rdfitc, the outflow persist to at least 7 Mm below the surface.

This agrees with the proposed picture described by \cite{2010ApJ...708..304Z} 
and others \citep{2009ApJ...698.1749H,2003Natur.421...43G}
with a circular flow below the sunspot penumbra with inflows at smaller depth and outflows deeper, forming a donut shape around 
the sunspot umbra.
The circular flow is believed to be powered by 
a downward flow caused by surface cooling within the plage which drags fluid in at shallow layers
\citep{2009ApJ...698.1749H}
and by the obstruction of the umbra pushing away the flow at deeper layers 
\citep[][and references within]{2013SoPh..287....9G, 2015SSRv..196..167K}.
Our results agree in general with estimations of individual sunspot 
by different authors
and are for a quiet region far away from the sunspot center.
\cite{2011JPhCS.271a2002F} 
observed outflows from just below the surface until 7 Mm deep for a sunspot observed by MDI in January 2002
using ring diagram analysis.
They estimated maximum flows around a depth of 5 Mm, 
which are 100 m~s$^{-1}$ at 28 Mm away from the sunspot's center.
\cite{2009SSRv..144..249G} observed a maximum outflow 30 Mm away from the center of sunspot NOAA 9787 at all depths from 1 to 4.5 Mm.
Around a depth of 4 Mm, the outflow has a maximum of 350 m~s$^{-1}$ and,
decreasing as the distance from the sunspot center increases, it is 20 m~s$^{-1}$ when 80 Mm away.
However, the estimated shallow inflow are of the same order (or much smaller, in the case of rdfitf)
than the outflow at deeper denser layers making the outward mass flux almost two orders of magnitude 
larger than inward.
Since we are observing here only a small slice of the proposed circulation flow, 
a more complete picture might be able to account for the mass conservation.

\section{Summary}

We analysed differences in the acoustic mode parameters at solar magnetically quiet regions (five-degree in diameter)
when there is an active region in their vicinity 
not farther than eight degrees away from the center of the quiet region,
with those of quiet regions at the same disc location for which there are no neighboring active region.
For each mode, 
these changes, dubbed nearby variations,
are averaged over all data obtained by the HMI Ring-Diagram processing pipeline from 2010 May to 2015 January.
We compared them with the variations observed in active regions 
with a mean line-of-sight magnetic field (MAI) between $110-120$ G
(in relation to quiet regions at the same location),
which we refer to as active region variations.
A summary of our results for each mode parameter is given below followed by the results for the horizontal flows.


{\bf Amplitude.}
The amplitude nearby variation changes from attenuation to enhancement at 4.2 mHz
which is at a much lower frequency than the active region variation, around 5.1 mHz.
The power reduction has a strong dependence on the wave direction.
The amplitude decrease by as much as 16\% if it is propagating in the direction of the nearby active region 
and only by less than 5\% if it is perpendicular to it.
The enhancement in power (known as acoustic halo effect) is as large as observed at an active region (almost 20\% at 5.3 mHz). 
There is some indication that the peak frequency is near 5.4 mHz.
We do not observe any significant variation in the $f$-mode amplitudes above 4.2 mHz.
There is a change in sign in the anisotropic component of the amplitude variation at 4.6 mHz.
Below 4.6 mHz, 
it is negative (waves are attenuated) in the direction of the nearby active region and
positive (they are amplified) perpendicular to it, and vice versa above 4.6 mHz.
The amplitude anisotropy is important below 4.6 mHz, it can vary by as much as 13\%
with the wave propagation direction. It is small at higher frequencies, 
varying only 3\% or less.
At a given frequency, the anisotropy decreases (in absolute value) as the mode degree increases.
The high inclined magnetic field in the vicinity of an active region might be responsible for
the smaller cutoff frequency and enhanced power \citep{2015LRSP...12....6K, 2015ApJ...801...27R, 2016ApJ...817...45R}.

{\bf Linewidth.} 
The change in sign in the linewidth nearby variation also happens at 4.2 mHz
and at a lower frequency than the active region variation (which is around 4.8 mHz).
The linewidth variation has an almost linear dependence with the amplitude variation.
For the active region variation, the slope is equal to $-1$, while for the nearby variation is $-2.6$.
The increase in the linewidth is much smaller (2\% or less), but the decrease is almost as large as in the active region (by as much as 9\%).

{\bf Energy.}
As for the mode amplitude and linewidth, the mode energy nearby variation changes sign at 4.2 mHz.
Below 4.2 mHz, 
the mode energy becomes smaller as the frequency decreases (until a given frequency) by as much as 8\%.
Otherwise, it increases with frequency.
As a result, the nearby mode energy variation averages to zero over all modes.
In active regions, there is a mode energy deficit at all frequencies,
being 50\% smaller on average for frequencies below 4.5 mHz.

{\bf Energy Rate.}
The temporal rate at which energy is pumped into or out of the mode is reduced by $-(40 \pm 10)$\% 
in active regions for all modes (assuming $\dot{E}$ is proportional to $A_c \, \gamma_c^2$),
with no frequency variation.
The energy rate variation in quiet nearby regions decreases with increasing
frequency, but does not usually change sign, merely reaching a maximum
of zero at around 4.5 mHz.

%
There is a distinct behavior of $p_4$ modes with respect to the other modes 
in linewidth nearby variation and, as consequence, in the energy and energy rate variations
and it
can not be easily
explained by a bias in the fitting methods or in the power spectra calculation since
their variation in the control set is small.

{\bf Asymmetry.}
Averaging over all modes with $n < 4$ in a quiet region, we get $S = -0.07$, which means that there is
7\% more power at the low-frequency side of the power spectrum peak than on the high-frequency one.
The asymmetry is larger (for most modes) at active regions.
The nearby asymmetry variation for the $f$ modes has a similar behavior as the active region variation scaled down by about 15 times. It is maximum around 2.8 mHz and decreases as frequency increases.
The $p$ modes have a different behavior, specially at high frequencies,
where, instead of a maximum increase in the asymmetry as in the active region (around 5.2 mHz), the nearby variation shows a maximum decrease in the asymmetry.

{\bf Frequency.}
Comparing mode frequency variations, we find similar behaviour in the
nearby regions as in active regions, with positive shifts
increasing slightly with increasing frequency until a point at which the
frequency shifts turn over sharply and become negative. 
The maximum increase is ten times smaller for the nearby variations.
The point at
which this turnover occurs in the quiet nearby regions, however,
is around 4.9 mHz, significantly below the acoustic cutoff frequency,
which is where it occurs in active regions. 
This could in principle be explained by a stronger horizontal magnetic field and an increase in temperature at the chrosmophere 
accordingly to a simple model by \cite{1996ApJ...456..399J}.
Below the turnover, the
frequency variations vary inversely with the mode inertia, implying that
their source is very close to the solar surface.
This is more so for the nearby variation, since the
linear regression of
the relative frequency variation multiplied by the scaled mode inertia (known as the surface term)
gives a standard deviation forty times smaller for the nearby regions than for the active regions.
The linear regression slope of the nearby surface term is
0.43 $\pm$ 0.03 mHz$^{-1}$ for rdfitc and 1.4 $\pm$ 0.1 mHz$^{-1}$ for rdfitf
in the frequency interval $2.4-4.9$ mHz.
%

{\bf Flow.}
The horizontal flow parameters in the direction of the nearby active regions were inverted
giving 
inflows toward the active regions at depths of around
1 Mm, and outflows below 4 Mm, with no net horizontal flow between. 
These results agree with the proposed picture of a circular flow below the sunspot penumbra, 
forming a donut shape below and around the sunspot umbra.


\acknowledgements

This research was supported in part by FAPEMIG and CNPq.



\facility{SDO (HMI)}


\bibliographystyle{aasjournal}

\bibliography{rabellosoares_bib}



\listofchanges

\end{document}